\documentclass[conference]{IEEEtran}
\usepackage{stfloats}
\usepackage{xcolor}
\usepackage{colortbl}
\usepackage[switch]{lineno}

\usepackage[fleqn]{amsmath} 
\usepackage{amsthm}
\theoremstyle{plain}
\newtheorem{lemma}{Lemma}
\newtheorem{theorem}[lemma]{Theorem}

\newtheorem{corollary}[lemma]{Corollary}
\theoremstyle{definition}

\colorlet{shadecolor}{gray!40}
\definecolor{apricot}{rgb}{0.98, 0.81, 0.69}

\makeatletter
\newcommand\notsotiny{\@setfontsize\notsotiny{6}{7}}

\makeatother

\usepackage{caption}
\usepackage{subfig}

\usepackage{url}
\usepackage{setspace}
\usepackage{placeins}
\usepackage{cite}

\usepackage{algorithm}
\usepackage[noend]{algpseudocode}
\algnewcommand{\LineComment}[1]{\State \emph{\textcolor{blue}{\(\triangleright\) #1}}}
\algrenewcommand\algorithmicindent{1em}%
\newcommand{\pluseq}{\mathrel{+}=}
\usepackage{graphicx}
\usepackage{tikz}
\usepackage{tabularx}
\usepackage{booktabs}
\usepackage{enumitem}   

\usepackage{color, soul} 
\soulregister\emph7
\soulregister\cite7
\soulregister\ref7
\soulregister\pageref7

\definecolor{Gray}{gray}{0.85}
\definecolor{LightCyan}{rgb}{0.88,1,1}

\usepackage{listings}

\newfloat{lstfloat}{htbp}{lop}
\floatname{lstfloat}{Listing}

\definecolor{mGreen}{rgb}{0,0.6,0}
\definecolor{mGray}{rgb}{0.5,0.5,0.5}
\definecolor{mPurple}{rgb}{0.58,0,0.82}
\definecolor{backgroundColour}{rgb}{0.95,0.95,0.92}

\lstdefinestyle{CStyle}{
    backgroundcolor=\color{backgroundColour},   
    commentstyle=\color{mGreen},
    keywordstyle=\color{magenta},
    numberstyle=\tiny\color{mGray},
    stringstyle=\color{mPurple},
    basicstyle=\notsotiny,
    breakatwhitespace=false,         
    breaklines=true,                 
    captionpos=b,                    
    keepspaces=true,                 
    numbers=left,                    
    numbersep=5pt,                  
    showspaces=false,                
    showstringspaces=false,
    showtabs=false,
    tabsize=2,
    language=C,
    moredelim=**[is][\color{blue}]{~}{~},
}

\newcolumntype{R}{>{\centering\arraybackslash}m{3.5cm}}
\newcolumntype{L}{>{\centering\arraybackslash}m{1.5cm}}
\newcolumntype{M}{>{\centering\arraybackslash}m{3.5cm}}

\begin{document}

\title{\fontsize{20.5pt}{20.5pt}\selectfont Space Filling Curves is All You Need: Communication-Avoiding Matrix Multiplication Made Simple}
\author{\IEEEauthorblockN{Evangelos Georganas, Alexander Heinecke, Pradeep Dubey}
\IEEEauthorblockA{Intel Corporation}}
\maketitle
\begin{abstract}
General Matrix Multiplication (GEMM) is the cornerstone of HPC workloads and Deep Learning. State-of-the-art (SOTA) vendor libraries tune tensor layouts, parallelization schemes and cache blocking to minimize data movement across the memory hierarchy and maximize throughput. However, optimal settings for these parameters depend on the target platform and matrix shapes, making exhaustive tuning infeasible. In this work, we address this cumbersome scheduling search and tuning using space-filling curves (SFC). We partition the matrix multiplication using advancements in SFC, and obtain platform-oblivious and shape-oblivious matrix multiplication schemes with a high degree of data locality. We extend the SFC-based work partitioning to implement Communication-Avoiding (CA) algorithms with replication techniques in a seamless fashion. The resulting SFC-CA GEMM achieves provable asymptotic communication optimality for both square and rectangular matrix regimes. Across four x86 and Arm platforms, SFC-CA GEMM outperforms vendor libraries by up to 5.5$\times$ per shape and 1.8$\times$ in weighted harmonic mean (WHM) throughput. Last, we show the impact of our work on two real-world applications by leveraging our SFC-CA GEMM as a compute backend: i) prefill of LLM inference with speedups up to 1.85$\times$ over SOTA inference runtimes, and ii) distributed-memory matrix multiplication with speedups up to 2.3$\times$ over the SOTA distributed-memory GEMM framework with vendor-optimized compute backend.
\end{abstract}

\section{Introduction}
\label{sec:introduction}
\begin{figure}[t]
\centering
\includegraphics[width= \columnwidth,trim=1.2bp 1.9bp 0.5bp 1.3bp,clip]{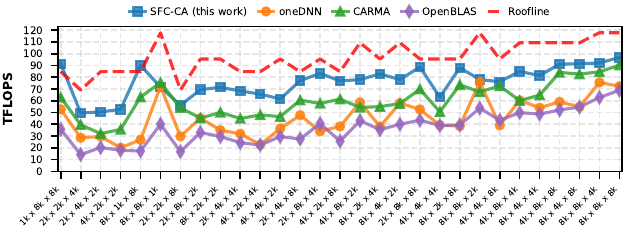}
\caption{BF16 GEMM performance on 128-core GNR: SFC-CA GEMM (this work), oneDNN, CARMA and OpenBLAS. The x-axis shows $M\times N\times K$ dimensions of GEMM.}
\label{fig:motivation}
\end{figure}

General Matrix Multiplication (GEMM) is the cornerstone of tensor contractions that are ubiquitous in High Performance Computing (HPC) and Deep Learning (DL) workloads, and academia and industry have therefore heavily optimized this kernel~\cite{meyer2023matrix,choi1996design,ben2019demystifying,di2022high,kepner2016mathematical,georganas2021tensor}. Modern platforms with matrix multiplication accelerators exhibit high FLOP/byte machine balance, making optimal matrix multiplication implementation a challenge~\cite{georganas2024harnessing}. For CPU platforms with matrix engines, state-of-the-art (SOTA) vendor libraries optimize the input tensor layouts, the parallelization schemes and the cache blocking to minimize data movement across the memory hierarchy and maximize throughput. However, the optimal settings of these parameters depend on the platform at hand (core-count, memory hierarchy, cache-sizes) and the shapes of the involved matrices, making optimal tuning of these libraries infeasible and leading to performance ``glass jaws". Figure~\ref{fig:motivation} shows (orange) the Bfloat16 (BF16) multi-core GEMM performance of the vendor-optimized oneDNN library on a 128-core Intel Xeon Granite Rapids (GNR) with Advanced Matrix Extensions (AMX), which exhibits such ``glass jaws" and is far off the roofline even for shapes with large operational intensity.

To tackle this tuning on CPUs with matrix accelerators, PARLOOPER~\cite{georganas2024harnessing} develops portable GEMM kernels from 1) a computational core of Tensor Processing Primitives (TPPs)~\cite{georganas2021tensor}, a compact set of 2D-tensor operators, and 2) logical loops around the TPPs, declared in a high-level fashion, whose instantiation (ordering, tiling, parallelization) is determined via simple knobs. Nevertheless, finding the knob value that maximizes performance for the platform and GEMM at hand requires time-consuming, platform-specific auto-tuning. Communication-avoiding GEMM algorithms instead exploit extra memory to reduce data movement~\cite{solomonik2011communication,georganas2012communication,demmel2013communication}. For shared memory, CARMA~\cite{demmel2013communication} recursively splits the largest dimension and achieves optimal asymptotic communication bounds and SOTA performance on older CPU architectures. However, its processor-halving property requires power-of-two core counts, which is a practical constraint~\cite{kwasniewski2019red}.

In this work we revisit space-filling curves (SFC)~\cite{bader2012space} to alleviate the cumbersome GEMM scheduling/tiling search. SFCs convert multi-dimensional coordinates (e.g.\ 2D) into a single dimension (1D), keeping nearby points in the high-dimensional space close in the 1D order. We partition the matrix multiplication computation space using recent advancements in generalized space-filling curves (generalized Hilbert curves~\cite{gilbert}), and we obtain platform-oblivious and shape-oblivious matrix multiplication schemes that exhibit an inherently high degree of data locality. Unlike previous work~\cite{heinecke2008parallel,chatterjee1999recursive} that used SFC to determine locality-friendly tensor layouts, but failed to generalize, yielded complicated indexing in the microkernels and offered limited performance upside over vendor-optimized libraries, we merely use the SFC to partition the \emph{tiled} GEMM computation space and we use the Batch-Reduce GEMM (BRGEMM) TPP for optimal single-core/single-output tensor-tile code generation. Furthermore, we extend the SFC-based work partitioning to implement Communication-Avoiding (CA) 2.5D GEMM algorithms~\cite{mccoll1999memory,solomonik2011communication, georganas2012communication} that replicate the tensors, and we prove that they minimize communication/data movement on the critical path of execution for both square and rectangular matrix regimes. The integration of the CA-algorithms with the generalized SFC-based partitioning is seamless and yields compact code ($\sim$30 LOC) with SOTA results on x86 and Arm CPUs (blue line in Figure~\ref{fig:motivation}). Due to its platform-oblivious and shape-oblivious nature, SFC-CA GEMM merely exposes a couple of algorithmic knobs, for which we derive simple analytical models, obviating the need for expensive auto-tuning. The contributions of this work are:

 \begin{itemize}[labelindent=0em,labelsep=0.2cm,leftmargin=*,noitemsep,topsep=0pt]
 \item A novel, SFC-based, Communication-Avoiding (SFC-CA) matrix multiplication algorithm that is inherently platform-oblivious and GEMM-shape-oblivious, and minimizes communication (i.e.\ data movement from memory/last-level cache to private L2 cache) along the critical path of execution. We provide proofs of asymptotic communication optimality for square GEMM and rectangular regimes.
\item Detailed GEMM benchmarking across various contemporary CPU platforms (x86 and Arm), illustrating portable SOTA performance that outperforms optimized vendor libraries up to 5.5$\times$ (up to 1.8$\times$ speedup considering Weighted Harmonic Mean (WHM) performance across a range of shapes). We analyze the performance by devising tight roofline models, and we develop methods to configure SFC-CA GEMM via analytical and ML performance models. Compared to the previous SOTA CA GEMM algorithm (CARMA), our work outperforms it by up to 3$\times$ on specific GEMM shapes and up to 1.27$\times$ considering WHM.
 \item We exhibit impact on 2 applications by integrating SFC-CA GEMM as a compute backend in: i) a SOTA CPU LLM inference framework (PyTorch-TPP~\cite{georganas2021tensor}) where it accelerates the compute-heavy prefill of LLM inference up to 1.85$\times$, ii) a SOTA distributed-memory matrix multiplication framework (COSMA~\cite{kwasniewski2019red}) where we observe speedups up to 2.3$\times$.
 \end{itemize}

\section{SFC-based Communication-Avoiding GEMM}
\label{sec:sfc}
\subsection{Tensor Processing Primitives and GEMM notations}
\label{subsec:gemm_prelim}
The TPPs~\cite{georganas2021tensor} constitute a set of single-threaded microkernels operating on 2D tensor tiles. In principle, given an architecture and tile shape, one can devise an optimal implementation for the TPPs that adopts conventional vectorization and register blocking optimizations. These single-threaded optimizations for the underlying microkernels are abstracted by the TPP layer and are undertaken by the TPP backend implementation which is platform-specific~\cite{georganas2021tensor}. We leverage the following TPPs as building blocks in this work: \emph{zero\_tpp}, \emph{add\_reduce\_tpp} and \emph{brgemm\_tpp}. The \emph{zero\_tpp} merely sets the input 2D tensor tile to zero. The \emph{add\_reduce\_tpp} accumulates multiple 2D tensor tiles that are separated by a fixed stride. The \emph{brgemm\_tpp} corresponds to the BRGEMM TPP, which is the main building block for general tensor contractions in the TPP collection~\cite{georganas2020harnessing,georganas2021tensor}. BRGEMM materializes the operation $C \pluseq \sum_{i=0}^{brcount-1} A_i \times B_i$. It multiplies the specified blocks $A_i^{bm\times bk}$ and $B_i^{bk\times bn}$ and reduces the partial results to a block $C^{bm\times bn}$. We use the \emph{stride-based} BRGEMM, where the addresses of $A_i$ and $B_i$ are: $address\_A_i = address\_A_{i-1} + stride\_A$ and $address\_B_i = address\_B_{i-1} + stride\_B$~\cite{georganas2021tensor}. The GEMM inputs are 2D matrices $A^{M\times K}$ and $B^{K\times N}$ that need to be multiplied and added onto $C^{M\times N}$. We follow the approach of prior work~\cite{georganas2020harnessing} and we block $M$, $K$, and $N$ by factors $bm$, $bk$, and $bn$ respectively (lines 1-3 of Listing~\ref{lst:gemm}). We employ the BRGEMM TPP for the tensor contraction with $A$ and $B$ across their dimensions $Kb$ and $bk$ (which form the $K$/inner-product dimensions of the original 2D matrices). 

\subsection{Generalized Hilbert Space-Filling Curve (SFC)}
\label{subsec:gilbert}
\begin{figure}[t]
\centering
\includegraphics[width=\columnwidth]{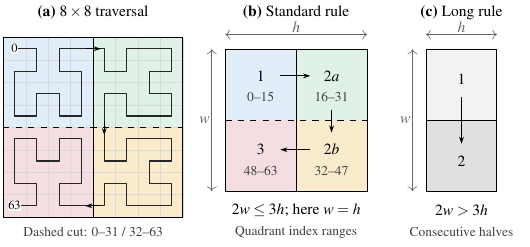}
\caption{Gilbert traversal and recursion~\cite{gilbert}. \textbf{(a)} An $8\times8$ traversal, with endpoints 0 and 63. \textbf{(b)} Standard rule on a square: matching colors identify the four consecutive quadrants and their index ranges in (a), and the dashed cut separates the two half-square runs, 0--31 and 32--63. \textbf{(c)} Long rule: visit the two halves consecutively. The major side $w$ is vertical and $h$ horizontal.}
\label{fig:hilbert-recursion}
\end{figure}

The discrete Hilbert curve maps multidimensional grid coordinates to 1D indices while preserving locality. Its square construction recursively visits four quadrants in a ``U" order, with rotations or reflections inside each quadrant~\cite{bader2012space}. The generalized Hilbert (\emph{Gilbert}) algorithm extends this construction to arbitrary rectangular grids, where $w$ is the local major side and $h$ the perpendicular side. Figure~\ref{fig:hilbert-recursion}(b,c) summarizes the recursion:
\begin{description}[leftmargin=0em,labelindent=0em,noitemsep,topsep=2pt]
\item[Base case.] Visit linearly a rectangle with a side of length 1.
\item[Long rule ($2w>3h$).] Visit the two halves of size $(w/2)\times h$ consecutively, starting with the half containing the starting corner (Figure~\ref{fig:hilbert-recursion}(c)). Recurse on each with the same major direction.
\item[Standard rule ($2w\le3h$).] Visit three parts in order: the $(w/2)\times(h/2)$ starting-corner block with directions swapped, then the $w\times(h/2)$ far-side strip, and finally the remaining corner block in reverse orientation (Figure~\ref{fig:hilbert-recursion}(b)). Recurse within each part.
\end{description}
For the square in Figure~\ref{fig:hilbert-recursion}(b), the right rectangular strip splits into $2a$ and $2b$, yielding the quadrant order $1,2a,2b,3$. The first two quadrants form one half of the square and the last two the other half of the square.

Figure~\ref{fig:hilbert-recursion}(a) illustrates the actual traversal on an $8\times8$ grid where the endpoint labels mark its first and last indices, and the quadrant colors match the index ranges in (b). Two observations inspire our SFC-CA GEMM algorithm (Section~\ref{subsec:sfc_algo}). First, consecutive indices in the grid visit edge-neighboring boxes (i.e.\ spatial locality). Second, recursive index ranges form locality-aware patches: indices 0--31 cover the top $4\times8$ half, indices 0--15 its top-left $4\times4$ quadrant, and indices 8--15 a $4\times2$ sub-patch.

\subsection{2.5D Communication-Avoiding GEMM Algorithm}
\label{subsec:25D_algo}
The 2.5D Communication-Avoiding GEMM~\cite{mccoll1999memory,solomonik2011communication} extends the classic 2D algorithms (Cannon's~\cite{cannon1969cellular}/SUMMA~\cite{van1997summa}) by replicating data. For $C = A \times B$ with $n \times n$ matrices, the $P$ processors form a $\sqrt{P/c} \times \sqrt{P/c} \times c$ grid, where $c$ is the replication factor ($1 \le c \le P^{1/3}$). Each of the $c$ layers holds a copy of $C$ and performs $1/c$ of the work on distinct $A$ and $B$ panels with logical inner-product dimension $n/c$. A final reduction across layers aggregates the partial results. With $O(cn^2/P)$ memory per processor, it moves $W = O(n^2/\sqrt{cP})$ words per processor, a factor $\sqrt{c}$ below the cost of 2D algorithms ($c=1$), trading memory for communication. In shared memory a ``message" is a cache line, so we focus on the number of words, since minimizing words on the critical path also minimizes the cache lines moved.

\subsection{SFC-Based Communication-Avoiding GEMM Algorithm}
\label{subsec:sfc_algo}
\begin{figure*}[!t]
\centering
\includegraphics[width=2\columnwidth]{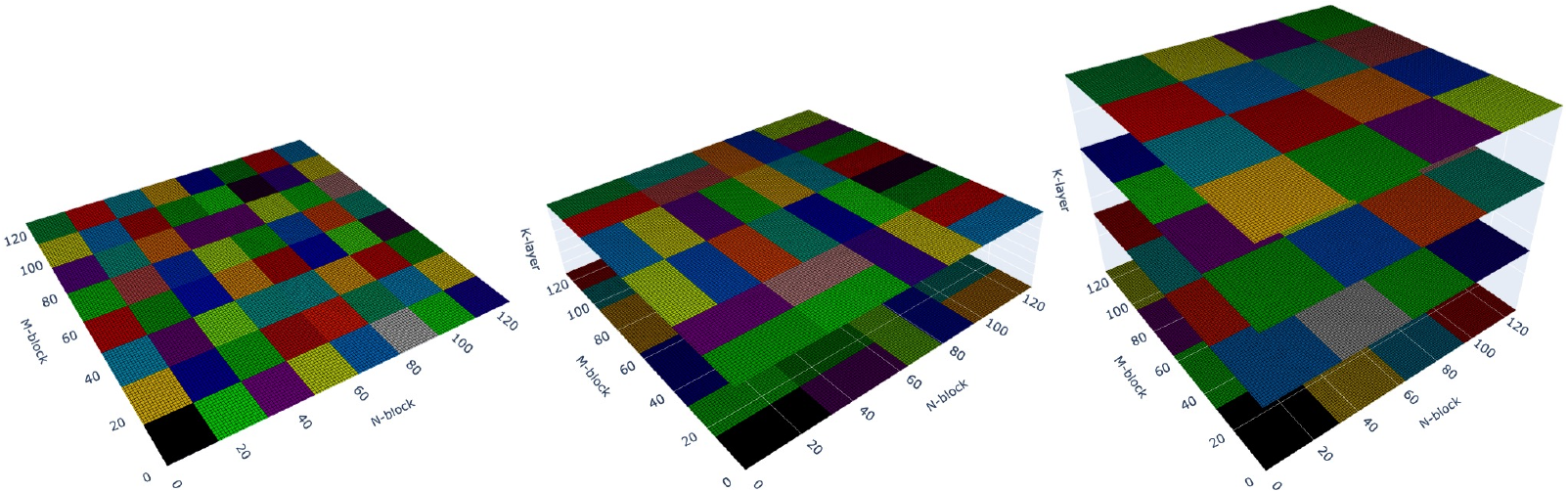}
\caption{SFC-based partitioning of a 4096$\times$4096 $C$ matrix (128$\times$128 blocks of size 32$\times$32) among 64 cores, one color per core. \textbf{Left:} 2D ($K\_layers=1$). \textbf{Middle:} 2.5D ($K\_layers=2$). \textbf{Right:} 3D ($K\_layers=4$).}
\label{fig:25_algo}
\end{figure*}

\begin{lstfloat}[t]
\begin{lstlisting}[style=CStyle]
DType A[Mb][Kb][bk][bm]; //Block M with bm and K with bk
DType B[Nb][Kb][bn][bk]; //Block N with bn and K with bk
DType C[K_layers][Nb][Mb][bn][bm]; //Block N with bn and M with bm
sfc_map = create_gilbert_sfc_map(Mb, Nb);
Kb_per_layer = Kb / K_layers;
Kb_per_brgemm = Kb_per_layer / k_block_factor;
for (ik = 0; ik < k_block_factor; ik++) {//K-block for A/B panel-size
#pragma omp parallel for schedule(static)
   for (i = 0; i < Mb * Nb * K_layers; i++) {
      i_layer = i / (Mb * Nb); // Index for K_layer 
      i_sfc = i % (Mb * Nb);   // 1D index for SFC within K-layer
      (im,in) = map_sfc_index(sfc_map,i_sfc);//1D SFC index -> im,in
      k_block_index = i_layer * Kb_per_layer + ik * Kb_per_brgemm; 
      if (ik == 0) zero_tpp(&C[i_layer][in][im][0][0]);//Zero C block
      brgemm_tpp(&A[im][k_block_index][0][0],
                 &B[in][k_block_index][0][0],
                 &C[i_layer][in][im][0][0], &Kb_per_brgemm);
   }
}
if (K_layers > 1) {// Reduce C matrices across K layers
#pragma omp parallel for schedule(static)
   for (i = 0; i < Mb * Nb; i++) {
      in = i % Nb;  im = i / Nb;
      reduce_stride = M * N;
      C_block_final = &C[0][in][im][0][0];
      add_reduce_tpp(C_block_final, K_layers, reduce_stride);
   }
}
\end{lstlisting}
\caption{SFC-based Communication-Avoiding GEMM}
\label{lst:gemm}
\end{lstfloat}
Listing~\ref{lst:gemm} is the pseudocode of our SFC-Based Communication-Avoiding shared-memory GEMM. We block $M$, $K$, $N$ by $bm$, $bk$, $bn$ (lines 1-3), so the computation space consists of $Mb\times Nb$ output $C$ tiles ($Mb = M/bm$, $Nb = N/bn$), and we create the Gilbert SFC of this 2D grid (line 4). The tunable parameter $K\_layers$ partitions the global $K$ dimension and equals the number of copies of $C$ (outer dimension of $C$ in line 3). Each $C$ copy is computed from the corresponding distinct $K$-partition of $A$ and $B$. The blocking factor $k\_block\_factor$ controls the $A$/$B$ sub-panel sizes within each $K$-partition (loop in line 7).

The heart of the algorithm is the loop in line 9: a ``task iteration space" of size $Mb\times Nb\times K\_layers$ is assigned in a block fashion to the $T$ threads by \texttt{\#pragma omp parallel for}. The modulo logic in lines 10-11 \emph{implicitly} maps the first $Mb\times Nb$ work-items (first $C$ layer) to the first ``team" of $T'=T/K\_layers$ cores, the second batch to the second team, etc. For $K\_layers = 1$ there is one $C$ copy and one team of $T$ cores. Within a team, the algorithm assigns contiguous 1D SFC indices from the range $0 \cdots Mb\times Nb - 1$ to each core. By leveraging the precomputed SFC map, we map each thread-specific $i\_sfc$ 1D SFC-index to the proper $im$ and $in$ block of the $C$ matrix copy/layer that this thread/core will be working on (line 12). Next, with a BRGEMM TPP we perform the corresponding contraction on the $im$ and $in$ block of the $C$ matrix copy by using the proper panels of $A$ and $B$ (lines 15-17). Finally, if we have $K\_layers > 1$, we perform a reduction among the $K\_layers$ of $C$ copies since each $C$ copy on each $K$-layer holds only a partially computed contraction (lines 20-28). 

This simple algorithm yields a class of GEMM decompositions shown in Figure~\ref{fig:25_algo}. For $K\_layers = 1$ we get a 2D GEMM decomposition, and for $K\_layers > 1$ we get effectively the Communication-Avoiding 2.5D and 3D GEMM decompositions~\cite{mccoll1999memory,solomonik2011communication}. Figure~\ref{fig:25_algo} uses $T=64$ cores and a 4096$\times$4096 $C$ matrix with 32$\times$32 blocks ($bm=bn=32$), i.e.\ a grid of 128$\times$128 $C$ blocks. For $K\_layers = 1$ (Left), the SFC \emph{implicitly} yields an 8$\times$8 core grid. Each core owns a patch of 16$\times$16 $C$ blocks that it traverses in SFC order, with inherent temporal and spatial locality, obviating the need for explicit cache blocking and loop reordering. For $K\_layers = 2$ (Middle), each of the two $C$ layers is partitioned among 32 cores with rectangular patches. For $K\_layers = 4$ (Right), each layer has a 4$\times$4 core grid, yielding a logical 4$\times$4$\times$4 3D grid of cores. All these decompositions arise \emph{implicitly} from the SFC properties and the loop-index mapping in lines 10-12.


\subsection{Communication optimality of SFC-CA GEMM Algorithm}
\label{subsec:sfc_optimal}
\begin{table*}[t]
\centering
\caption{Proved asymptotic communication-optimal regimes and coverage of the dimension/thread ranges.}
\label{tab:optimality}
\begingroup
\small
\setlength{\tabcolsep}{3pt}
\setlength{\aboverulesep}{1.5pt}
\setlength{\belowrulesep}{1.5pt}
\setlength{\smallskipamount}{1pt}
\renewcommand{\arraystretch}{1}
\begin{tabularx}{\textwidth}{@{}>{\raggedright\arraybackslash}p{0.25\textwidth}>{\raggedright\arraybackslash}X>{\raggedright\arraybackslash}p{0.23\textwidth}@{}}
\toprule
\textbf{Matrix shape / regime}
& \textbf{Dimension/thread conditions and layer choice}
& \textbf{SFC-CA GEMM words $W$}\newline Matched lower-bound type\\
\midrule
Square: 2D / 2.5D\newline $M=N=K=n$\enspace (Cor.~\ref{cor:optimality}(i))
& $1\le c\le T^{1/3}$, $bk=b$, and $FastMem=\Theta(cn^{2}/T)$, in addition to the shared cache-fit condition. $c=1$ gives 2D.
& $\Theta\!\big(n^{2}/\sqrt{cT}\big)$\newline Memory-dependent\\
\midrule[0.3pt]
1D output bands\newline Any $K$\enspace (Cor.~\ref{cor:optimality}(ii))
& $1\le T\le M/N$ with $c=1$. Each core owns $M/T$ full rows of $C$.
& $\Theta\!\big((MK+MN)/T\big)$\newline Compulsory input/output\\
\midrule[0.3pt]
Short inner dimension\newline $K\le N\le M$\enspace (Cor.~\ref{cor:optimality}(iii))
& $M/N\le T\le MN/K^{2}$ with $c=1$. Writing $C$ dominates the total traffic.
& $\Theta(MN/T)$\newline Compulsory output\\
\midrule[0.3pt]
Large inner dimension\newline $K>N$, $M\ge N$\enspace (Cor.~\ref{cor:optimality}(iv))
& $M/N\le T\le MK/N^{2}$. The exact choice $c=\min(K/N,T)$ must be feasible.
& $\Theta(MK/T)$\newline Compulsory input $A$\\
\midrule[0.3pt]
Three large dimensions\newline Any ordering of $K$\enspace (Cor.~\ref{cor:optimality}(v))
& $T\ge\max(MN/K^{2},MK/N^{2})$. Assume $bk=b$ and $T,K_b$ are powers of two. Choose $c=2^{\lfloor\log_2 c_*\rfloor}$ and $f=K_b/c$.
& $\Theta\!\big((MNK/T)^{2/3}\big)$\newline Memory-independent\\
\bottomrule
\end{tabularx}
\par\smallskip
\begin{minipage}{\textwidth}
\textbf{Notations:} Assume $M\ge N$ (transpose otherwise). $T$ is the total core count, $c=K\_layers$, $T'=T/c$, and $f=k\_block\_factor$. $W$ counts words on the busiest core, including reduction.
\par\smallskip
\par\smallskip
\textbf{Shared hypotheses.} Use $bm=bn=b$, with $M_b=M/b$, $N_b=N/b$ and $T'$ powers of two, $T\le M_bN_b$, and positive integers satisfying $T\bmod c=0$, $M_bN_b\bmod T'=0$, $K_b\bmod (cf)=0$ ($K_b=K/bk$).
\par\smallskip
\textbf{Cache fit (fully associative LRU).} Each actual $h\times w$ patch requires $FastMem\ge hwb^{2}+2(h+w)bK/(cf)$, sufficient for retaining the output patch across successive panel sets.
\end{minipage}
\endgroup
\end{table*}
We prove the asymptotic communication-optimal regimes in Table~\ref{tab:optimality} for classical GEMM without recomputation, using the notation and shared hypotheses of the table. The power-of-two assumptions only simplify the proofs and are \emph{not} a limitation of SFC-CA GEMM, which runs on arbitrary core counts (e.g.\ 96 cores in Section~\ref{subsec:rest_results}).
In this theoretical model, each core has a private, word-addressed, fully associative LRU cache with capacity of $FastMem$ words (we call this private cache \emph{fast memory}). Main memory and the shared last-level cache are \emph{slow memory}. Inputs are initially only in slow memory and the final output ends there. As in Listing~\ref{lst:gemm}, output is zero-initialized without reading its old value. All bounds are asymptotic with constants independent of $M,N,K,T,c$.

\subsubsection{The geometry of the patches}
\label{subsubsec:square_decomp}
Within each layer, each of the $T'=T/c$ cores receives $M_bN_b/T'$ consecutive tiles in Gilbert traversal order. We call this set of tiles the core's \emph{patch}. The next lemma characterizes these patches using the recursion of Section~\ref{subsec:gilbert}.

\begin{lemma}[Patch geometry]
\label{lem:patch}
Without loss of generality, assume $M_b\ge N_b$. Let $M_b=2^{\alpha}$, $N_b=2^{\beta}$ and $T'=2^e\le M_bN_b$, with nonnegative integer exponents. The Gilbert traversal is dyadic: either the grid is a single tile, or it can be bisected across a longest side into two congruent rectangular halves such that the curve visits one half completely before the other and its restriction to each half is again dyadic. Each such bisection halves the aspect ratio when it is at least $2$ and turns a square into a $2$:$1$ rectangle. It also preserves alignment: the starting row and column are multiples of the respective side lengths. Consequently, the equal block partition among $T'$ workers gives each worker an aligned $h\times w$ rectangular patch of $M_bN_b/T'$ tiles, obtained by $e$ successive longest-side bisections. With $\rho_0=M_b/N_b$, every patch's aspect ratio $\rho=\max(h,w)/\min(h,w)$ is
\[
\rho=
\begin{cases}
\rho_0/T' & T'\le\rho_0,\\
1 & T'>\rho_0,\ T'/\rho_0\text{ is a power of }4,\\
2 & T'>\rho_0,\ T'/\rho_0\text{ is not a power of }4.
\end{cases}
\]
The three cases give, respectively, full-width $(M_b/T')\times N_b$ bands, aligned squares, and halves of aligned squares. In particular, $\rho\le\max(2,\rho_0/T')$.
\end{lemma}

\begin{proof}
\emph{Dyadic recursion.} We induct on $\alpha+\beta$ for rectangles with major and perpendicular side lengths $2^\alpha\ge2^\beta$.
For $\alpha+\beta=0$ the grid is a single tile. If $\beta=0$ and $\alpha>0$, the curve traverses a line of $2^\alpha$ tiles. Its two halves are visited consecutively, each linearly on $2^{\alpha-1}$ tiles, so each is dyadic by induction on the smaller exponent $\alpha-1$.

If $\alpha>\beta\ge1$, then $2^\alpha\ge2\cdot2^\beta$, so the long rule (Figure~\ref{fig:hilbert-recursion}(c)) visits the two $2^{\alpha-1}\times2^\beta$ halves consecutively. Their major side remains a longest side because $\alpha-1\ge\beta$. Each recursive call operates on a power-of-two rectangle with exponent sum $\alpha+\beta-1$. By induction its traversal is dyadic, so the linear traversal of the two $2^{\alpha-1}\times2^\beta$ halves is dyadic too.

If $\alpha=\beta\ge1$, the standard rule visits the quadrants $1,2a,2b,3$ of Figure~\ref{fig:hilbert-recursion}(b). The first and last recursive calls operate on $2^{\alpha-1}\times2^{\alpha-1}$ squares, possibly rotated or reflected. The middle call operates on the $2^\alpha\times2^{\alpha-1}$ strip: the long rule splits it into quadrants $2a,2b$ consecutively, or the linear base case does so when $\alpha=1$. Every quadrant is therefore traversed dyadically by induction, since its exponent sum is $2\alpha-2<2\alpha$. The first pair $1,2a$ forms one half of the parent square and the last pair $2b,3$ forms the other. Within each half, its two quadrants are precisely its longest-side halves and are visited consecutively. Thus the traversal of these halves, and hence the traversal of the parent square, are dyadic.

\emph{Patches and alignment.} Each dyadic bisection replaces one consecutive index interval by two consecutive intervals of equal length. Repeating this $e$ times from the full grid produces $2^e=T'$ rectangles with intervals of length $M_bN_b/T'$, exactly the equal blocks assigned to the workers. The grid starts at $(0,0)$ and is aligned. If a parent side of length $d$ starts at $md$ for an integer $m$, its two halves start at $md=2m\frac d2$ and $md+\frac d2=(2m+1)\frac d2$. Both are multiples of the new side length $d/2$. The other side and its starting coordinate are unchanged. Thus every descendant patch remains aligned.

\emph{Shapes and aspect ratios.} Halving a longest side halves an aspect ratio of at least $2$, while bisecting a square produces aspect ratio $2$. For $e\le\alpha-\beta$, only the original row side is halved repeatedly $e$ times, giving $(M_b/T')\times N_b$ bands with ratio $\rho_0/T'$. Otherwise, when $e>\alpha-\beta$, the patches are squares after the first $\alpha-\beta$ bisections. The remaining $e-(\alpha-\beta)$ bisection cuts alternate between half-squares and squares, ending at a square exactly when this number is even. Since $T'/\rho_0=2^{e-(\alpha-\beta)}$, even parity is equivalent to $T'/\rho_0$ being a power of four. With odd parity the final cut bisects an aligned square, proving the half-square case (i.e.\ aspect ratio 2).
\end{proof}

Writing $s=\sqrt{M_bN_b/T'}$, aligned square patches form an $(M_b/s)\times(N_b/s)$ thread grid with the matrix's aspect ratio (Figures~\ref{fig:25_algo}-Left and~\ref{fig:rectangular}). Half-squares have sides $s/\sqrt2$ and $s\sqrt2$ (Figure~\ref{fig:25_algo}-Middle) since the square before the last bisection had side $s'=\sqrt{M_bN_b/(T'/2)}$, and the bisection halves one side.

\subsubsection{Communication cost analysis}
\label{subsubsec:patchcost}
The number of words moved from slow memory to fast memory during the GEMM phase (i.e.\ excluding potential reduction) depends on the area and half-perimeter of each patch. We derive communication costs and present the following bound, which applies to any rectangular patch satisfying the conditions of Lemma~\ref{lem:patch}. Our analysis also includes the potential cost of reduction.

\begin{theorem}
\label{thm:patch}
Suppose each core $t$ owns a rectangular patch of $h$ tile-rows and $w$ tile-columns (possibly varying with $t$), $hw=M_bN_b/T'$, and each core's fast memory has capacity:
\begin{equation}
\label{eq:lru-fit}
FastMem\ge hwb^{2}+2(h+w)bK/(cf).
\end{equation}
Let $W_t$ count core $t$'s total tensor-word transfers, including reduction, and let $W=\max_t W_t$. If core $t$ is assigned $r_t\le\lceil M_bN_b/T\rceil$ reduction tiles by the balanced streaming reduction, then
\begin{equation}
\label{eq:W}
W_t\le(h+w)bK/c+cMN/T+[c>1](c+1)b^{2}r_t.
\end{equation}
The indicator $[c>1]$ restricts the reduction term only when $K\_layers>1$.
If $\rho$ is the largest patch aspect ratio, then
\begin{equation}
\label{eq:Wtheta}
W=\Theta\!\Big((\sqrt\rho+\rho^{-1/2})K\sqrt{\tfrac{MN}{cT}}
+\tfrac{cMN}{T}+[c>1]cb^{2}\big\lceil\tfrac{M_bN_b}{T}\big\rceil\Big).
\end{equation}
For $T\le M_bN_b$ and $\rho=O(1)$ this simplifies to
\begin{equation}
\label{eq:W-balanced}
W=\Theta\!\Big(K\sqrt{\tfrac{MN}{cT}}+\tfrac{cMN}{T}\Big).
\end{equation}
\end{theorem}

\begin{proof}
\emph{Patch data and reuse.} The static work-chunk length is $L=M_bN_b/T'$. Since a layer contains $T'L$ tiles, layer boundaries are chunk boundaries: each core stays within one layer and contracts over $K/c$ inner-product indices, evaluating $hwb^2K/c=MNK/T$ multiply-add operations. In one outer iteration (i.e.\ one iteration of line 7 of Listing~\ref{lst:gemm}) its tasks share an $hb\times K/(cf)$ panel of $A$ and a $K/(cf)\times wb$ panel of $B$, totaling $(h+w)bK/(cf)$ words. Between uses of an input word, the distinct words accessed lie in this output patch and $A$/$B$ panel set, so~\eqref{eq:lru-fit} ensures one load per $A$/$B$ word given LRU cache policy. For a $C$ word across consecutive iterations the reuse distance is at most $hwb^2-1+2(h+w)bK/(cf)<FastMem$ (the rest of $C$ and two panel sets), so LRU retains it. The $f$ iterations use disjoint $K$-ranges, giving exactly $(h+w)bK/c$ input reads and at most $hwb^2=cMN/T$ output writes.

\emph{Total $A$ and $B$ data movement.} The input word-count $R_{AB,t}$ is $K/c$ times $(h+w)b$, half the patch's perimeter. Let $\rho_t=\max(h,w)/\min(h,w)$. Interchanging the sides if necessary, the area $hwb^2=cMN/T$ gives $hb=\sqrt{cMN\rho_t/T}$ and $wb=\sqrt{cMN/(T\rho_t)}$. Thus $R_{AB,t}=(\sqrt{\rho_t}+\rho_t^{-1/2})K\sqrt{MN/(cT)}$.
The factor increases for $\rho_t\ge1$ and equals its minimum $2$ at a square, so the largest input count uses $\rho=\max_t\rho_t$.

\emph{Reduction cost.} Balanced reduction assigns a floor or ceiling of $M_bN_b/T$ tiles per core. Streaming each entry's $c$ partials through a register accumulator costs at most $c$ reads and one write, proving~\eqref{eq:W}. For $c=1$, final output writes force $W\ge MN/T$. For $c>1$, a busiest reducer produced at most one partial per entry, so it must receive at least $(c-1)b^2\lceil M_bN_b/T\rceil$ words from other cores. This is $\Omega(cb^2\lceil M_bN_b/T\rceil)$ and dominates $cMN/T$. Together with the exact input count $R_{AB,t}$, these bounds establish~\eqref{eq:Wtheta}, since even if the largest costs occur on different cores, their maximum and sum differ only by a constant.

\emph{Simplification.} If $T\le M_bN_b$, then $M_bN_b/T\ge1$ and $\lceil M_bN_b/T\rceil\le M_bN_b/T+1\le2M_bN_b/T$.
Since $b^2M_bN_b=MN$, reduction is $O(cMN/T)$ and is absorbed into the output term. Bounded $\rho\ge1$ makes $\sqrt\rho+\rho^{-1/2}=\Theta(1)$, yielding~\eqref{eq:W-balanced}.
\end{proof}


\subsubsection{Communication-optimal regimes}
\label{subsubsec:rect}
\begin{figure*}[t]
\centering
\includegraphics[width=1.9\columnwidth]{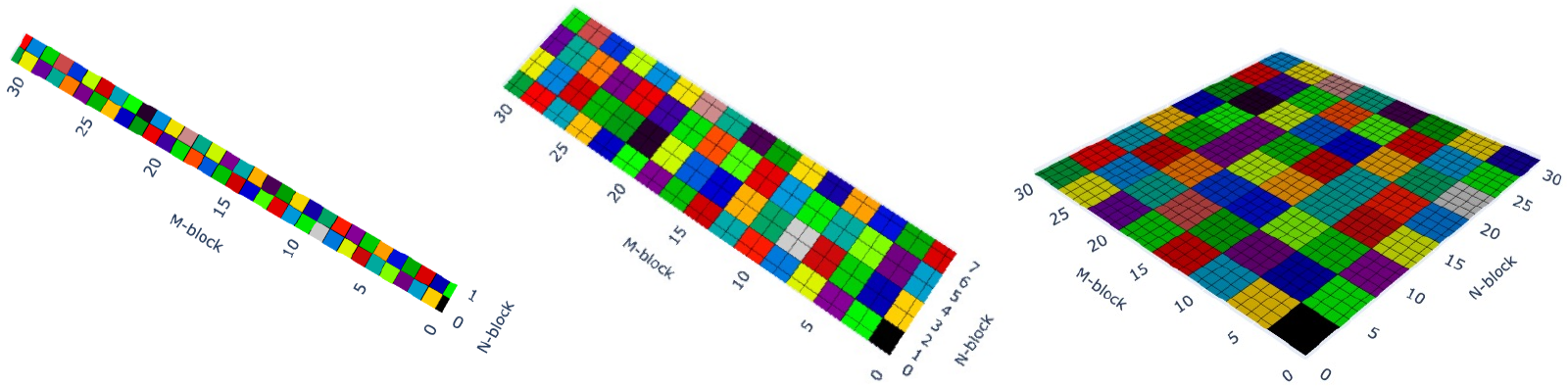}
\caption{SFC-based partitioning of $C$ matrices with different $M$:$N$ aspect ratios using 64 cores. The SFC-based partitioning yields 2D core decompositions with aspect ratios matching the corresponding $M$:$N$ ratio.}
\label{fig:rectangular}
\end{figure*}
Assume $M\ge N$ (transpose otherwise, Figure~\ref{fig:rectangular}). Cold inputs and final output in slow memory require (mandatory reads/writes) $W=\Omega((MK+KN+MN)/T)$. We also use the classical memory-dependent bound $\Omega(MNK/(T\sqrt{FastMem}))$ and two memory-independent bounds~\cite{demmel2013communication}: $\Omega(\sqrt{MNK^2/T})$ for $K\le N$ and $M/N\le T\le MN/K^2$ (two large dimensions), and $\Omega((MNK/T)^{2/3})$ for $T\ge\max(MN/K^2,MK/N^2)$ (three large dimensions).

\begin{corollary}[Communication optimality]
\label{cor:optimality}
Under the shared hypotheses of Table~\ref{tab:optimality}, SFC-CA GEMM is communication-optimal for (i) square matrices, (ii) 1D output bands, (iii) short inner dimension, (iv) large inner dimension and (v) three large dimensions, with the respective row conditions, layer choices and costs in that table.
\end{corollary}

\begin{proof}
\emph{(i) Square matrices.} Lemma~\ref{lem:patch} gives $\rho\le2$. Equation~\eqref{eq:W-balanced} yields $W=\Theta(n^2/\sqrt{cT}+cn^2/T)=\Theta(n^2/\sqrt{cT})$ because $c^3\le T$. The memory-dependent lower bound matches for $FastMem=\Theta(cn^2/T)$. This size is compatible with~\eqref{eq:lru-fit}: the panel term $2(h+w)bn/(cf)=\Theta(n^2/(f\sqrt{cT}))$ is $O(cn^2/T)$ for some $f=\Theta(\sqrt{T/c^3})$, which is feasible ($f\le K_b/c$) because $bk=b$ and $T\le M_bN_b$. An admissible $c=\Theta(T^{1/3})$ gives $W=\Theta(n^2/T^{2/3})$, matching the memory-independent bound.

\emph{(ii) 1D output bands.} With $c=1$ and $T\le M/N$, Lemma~\ref{lem:patch} gives scalar patches $(M/T)\times N$. Their input traffic is $MK/T+NK$, and their output traffic is $MN/T$. Since $NK\le MK/T$, the total is $\Theta((MK+MN)/T)$, matching the compulsory cost of reading $A$ and writing $C$, for any $K$.

\emph{(iii) Short inner dimension.} Here $c=1$ and $T\ge M/N$ ensure $\rho\le2$. The condition $T\le MN/K^2$ implies $K\sqrt{MN/T}\le MN/T$, so~\eqref{eq:W-balanced} gives $W=\Theta(MN/T)$, matching compulsory output. The input component $\Theta(K\sqrt{MN/T})$ also matches the two-large-dimensions bound~\cite{demmel2013communication}.

\emph{(iv) Large inner dimension.} If $K/N\le T$, choose $c=K/N$. Then $T'=TN/K\le M/N$, so patches are bands of scalar size $(MK/(TN))\times N$, contracting over $K/c=N$. Their input traffic is $MK/T+N^2$, and output and reduction add $O(cMN/T)=O(MK/T)$. Since $T\le MK/N^2$, also $N^2\le MK/T$, thus the total traffic is $\Theta(MK/T)$.

If instead $K/N>T$, choose $c=T$: each core owns an entire $M\times N$ layer, reads $(M+N)K/T\le2MK/T$ words, and incurs $O(MN)$ output/reduction traffic. Now $T<K/N$ implies $MN<MK/T$. Both branches attain $W=\Theta(MK/T)$, matching compulsory input $A$.

\emph{(v) Three large dimensions.} For this case only, additionally assume $bk=b$ and $T,K_b$ are powers of two. Set $c_*=(K^2T/(MN))^{1/3}$ and choose $c=2^{\lfloor\log_2 c_*\rfloor}$, so $c_*/2<c\le c_*$. The thresholds $T\ge MN/K^2$ and $T\ge MK/N^2$ give $c_*\ge1$ and $T/c_*\ge M/N$, respectively. Hence $c\le T$ and $T'=T/c\ge M/N$. Also, $T\le M_bN_b$ gives $c_*^3=K_b^2T/(M_bN_b)\le K_b^2$, so $c\le K_b$. Since $c,T,K_b$ are powers of two, $T\bmod c=K_b\bmod c=0$. Therefore $T'$ is a power of two and, since $T'\le M_bN_b$, each layer's tiles divide evenly among its cores. Choosing $f=K_b/c$ gives $K_b/(cf)=1$, so each panel is one tile deep along $K$.

Since $T'\ge M/N=M_b/N_b$, Lemma~\ref{lem:patch} gives square or half-square patches, with $\rho\le2$. Together with the assumed cache capacity~\eqref{eq:lru-fit}, this lets us apply the simplified traffic bound~\eqref{eq:W-balanced}. At $c=c_*$, its input and output/reduction terms are equal:
\[
K\sqrt{\frac{MN}{c_*T}}=\frac{c_*MN}{T}
=\left(\frac{MNK}{T}\right)^{2/3}.
\]
Replacing $c_*$ by $c$ multiplies the input term by $\sqrt{c_*/c}\in[1,\sqrt2)$ and the output/reduction term by $c/c_*\in(1/2,1]$. These are constant factors, so~\eqref{eq:W-balanced} gives $W=\Theta((MNK/T)^{2/3})$, matching the memory-independent lower bound~\cite{demmel2013communication}.
\end{proof}

\section{SFC-CA GEMM Implementation}
\subsection{Implementation details}
\label{subsec:sfc_impl_details}
We implemented the SFC-CA GEMM algorithm in C++ and the code is virtually identical to Listing~\ref{lst:gemm}. We used OpenMP to partition the 1D SFC index space in a block fashion. However, \emph{any} other block partitioning of the 1D SFC index space has the same effect on the efficacy of SFC-CA GEMM. For the TPP building blocks we used the implementation from LIBXSMM~\cite{heinecke2016libxsmm}, which offers optimized machine code generation Just-In-Time (JIT) for x86 and Arm/AArch64 architectures. For the generation of SFC index maps we leveraged the code for arbitrary rectangular domains~\cite{gilbert}. Last, the precision in the SFC-CA GEMM implementation of Listing~\ref{lst:gemm} may be parameterized, i.e.\ DType can be BF16, single precision, double precision, FP8, int8 etc. The implementation stays the same, and all the precision-specific code generation is undertaken by the TPP abstraction/backend, which emits at runtime (i.e.\ JIT) optimal machine code for the precision and machine at hand. 

\subsection{Performance modeling}
\label{subsec:sfc_perf_model}
To assess the efficiency of SFC-CA GEMM, we devise \emph{fine-grained} roofline models by applying the roofline model~\cite{williams2009roofline} at the BRGEMM granularity instead of the whole GEMM, which yields a \emph{tight} bound. Assuming infinite fast memory per core, a BRGEMM in SFC-CA GEMM is of type $BRGEMM_0$ ($A$ and $B$ panels from slow memory), $BRGEMM_1$ (only $A$ from slow memory, $B$ reused from a previous BRGEMM on the same core), $BRGEMM_2$ (only $B$ from slow memory) or $BRGEMM_3$ (both in fast memory). Given a GEMM shape and a thread decomposition, the BRGEMM types on the critical path are determined analytically. Let $\gamma$ be the inverse compute rate per core (cycles per flop) with operands in fast memory, $\beta$ the inverse read bandwidth of a core from slow memory (cycles per byte), $G$ the flops per BRGEMM, and $S_A$, $S_B$ the $A$/$B$ panel sizes in bytes. Assuming $C$ loads/stores are serviced from fast memory, the time $t_i$ (in cycles) of a $BRGEMM_i$ is:
\begin{align}
t_i&=\max\big(G\cdot\gamma,\ \beta\cdot S_i\big),\qquad i=0,\dots,3,\\
S_0&=S_A+S_B,\quad S_1=S_A,\quad S_2=S_B,\quad S_3=0.\nonumber
\end{align}
We sum the times of all BRGEMMs on the critical path and add the time to read/write $C$ from/to slow memory, including the $C$ copies of the final reduction for 2.5D/3D. For a given $M\times N\times K$ and core count, the model iterates over all 2D and 3D thread decompositions, with $\beta$ and $\gamma$ from microbenchmarks~\cite{heinecke2016libxsmm,archbench}, and reports the \emph{minimum} time (maximum throughput) as the roofline.

\subsection{Runtime configuration of SFC-CA GEMM}
\label{subsec:tuning}
As evident from Listing~\ref{lst:gemm}, the only tunable algorithmic knobs are the number of $C$ copies ($K\_layers$) and $k\_block\_factor$, which determines the $A$/$B$ panel sizes of the BRGEMM calls. We implemented 3 ways to pick them for the GEMM at hand:
\begin{enumerate}[labelindent=0em,labelsep=0.05cm,leftmargin=*,noitemsep,topsep=0pt]
\item \ul{A priori autotune the specific shape ($M$,$N$,$K$)} over the tuples ($K\_layers$, $k\_block\_factor$). The search space is small (e.g.\ $K\_layers \leq 8$ and $k\_block\_factor \leq 8$, i.e.\ 64 combinations per shape), so it can be benchmarked exhaustively.
\item \ul{Use the analytical model to predict ($K\_layers$, $k\_block\_factor$)}. The roofline model of Section~\ref{subsec:sfc_perf_model} recommends $K\_layers$. Since it assumes infinite per-core L2, we then adjust $k\_block\_factor$ such that the \emph{total} $A$ and $B$ panel footprint per thread (accessed in lines 9-18 of Listing~\ref{lst:gemm}) is within a fraction (e.g.\ $0.5$) of the L2 cache.
\item \ul{Use a Machine Learning (ML) model}. We frame the prediction as classification with a Nearest-Neighbor (NN) model: GEMM shapes that are close in the 3D ($M$,$N$,$K$) space likely share the optimal ($K\_layers$, $k\_block\_factor$). We auto-tuned 1573 GEMM shapes with $512 \leq M \leq 16384$, $128 \leq N \leq 16384$ and $512 \leq K \leq 16384$ covering a cuboid space. Given a new ($M'$,$N'$,$K'$), we use the tuple ($K\_layers$, $k\_block\_factor$) of its nearest neighbor.
\end{enumerate}

\section{Results}
\label{sec:results}
\begin{figure*}[t]
\centering
\includegraphics[width=\textwidth]{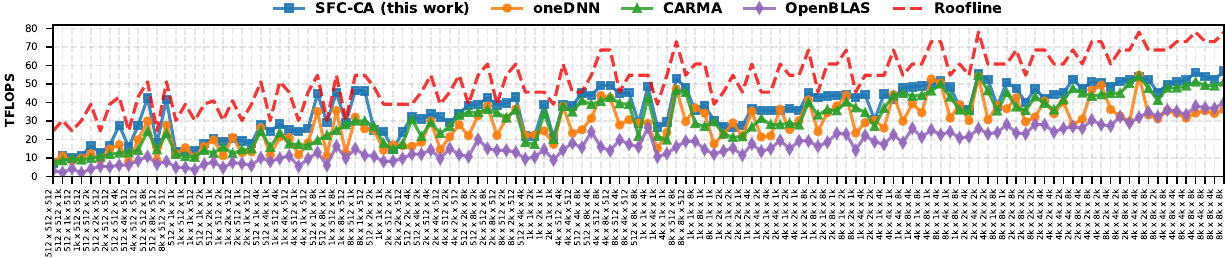}
\caption{EMR BF16 GEMM performance over all 125 shapes: SFC-CA GEMM, oneDNN, CARMA, OpenBLAS and the roofline. The $M\times N\times K$ configurations are sorted by operational intensity.}
\label{fig:perf_emr}
\end{figure*}

\subsection{Experimental Platforms and Setup}
\label{subsec:exp_platforms}
We experimented with contemporary CPU platforms, which offer various matrix multiplication acceleration engines:

\textbf{Emerald Rapids (EMR)}: An Intel Xeon Platinum 8592+ CPU with 64 cores and 512GB DDR5@5600 memory. EMR offers Intel AMX technology for matrix multiplication~\cite{intelisa}.

\textbf{Granite Rapids (GNR)}: An Intel Xeon 6980 CPU with 128 cores, 768GB DDR5@8800 and AMX technology.

\textbf{5th Gen AMD EPYC (ZEN5)}: A 5th Gen AMD EPYC with 96 cores (AWS bare-metal m8a.metal-24xl instance). BF16 multiply-add is accelerated via AVX512 BF16 FMAs.

\textbf{Graviton-4 (GVT4)}: A 4th Gen AWS Graviton with 96 cores (bare-metal r8g.metal-24xl instance). The GVT4 Instruction Set is based on Armv9.0-A, featuring Neoverse-V2 cores. BF16 matrix multiplication is boosted via BFMMLA instructions~\cite{georganas2024harnessing}.

We use 125 matrix shapes (cross product of $M$, $N$, $K$ from $\{512, 1024, 2048, 4096, 8192\}$) with various aspect ratios of $A$, $B$, $C$, in BF16~\cite{bfloat16_tf} to leverage the acceleration instructions. Baselines include oneDNN~\cite{onednn} on x86 (EMR/GNR/ZEN5) and the optimized Arm Compute Library (ACL)~\cite{acl} via the oneDNN frontend on GVT4, both with library-selected layouts (format tag \emph{any}). On EMR and GNR we also compare against CARMA and OpenBLAS~\cite{OpenBLAS}. We ported the CARMA implementation~\cite{demmel2013communication} to use the same blocked tensor layouts and TPPs as SFC-CA GEMM. This keeps the compute kernels and tensor layouts fixed, so the remaining performance differences can be attributed to schedule and data-movement characteristics.

We benchmark sets of $A$/$B$/$C$ matrices large enough that inputs come from memory (``cold" tensors), repeat each experiment 10 times and report throughput as $2MNK$ divided by the mean time (run-to-run variation below 2\%). For SFC-CA GEMM we try all $K\_layers, k\_block\_factor\in\{1,\dots,8\}$ and report the best. Per-platform summaries use the WHM~\cite{smith1988characterizing,hoefler2015scientific} of throughput over the 125 shapes, weighted by operation count. For shape $i$, with $F_i=2M_iN_iK_i$ operations, time $t_i$ and throughput $P_i=F_i/t_i$, this is $P_{\rm WHM}=(\sum_i F_i)/(\sum_i F_i/P_i)=(\sum_i F_i)/(\sum_i t_i)$. The roofline parameters $\beta$ and $\gamma$ come from LIBXSMM~\cite{heinecke2016libxsmm} and ArchBench~\cite{archbench} microbenchmarks. Finally, we show the impact of SFC-CA GEMM as a compute backend on LLM inference prefill and on distributed-memory matrix multiplication.

\subsection{Experimental results on EMR}
\label{subsec:emr_results}
\begin{figure}[t]
\centering
\includegraphics[width=\columnwidth]{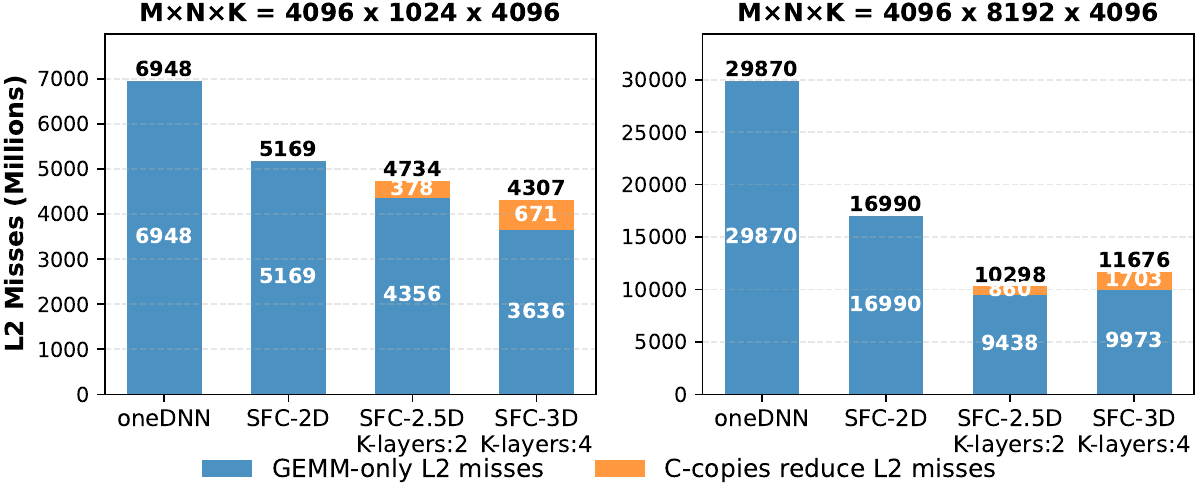}
\caption{Total L2 misses on EMR for two problem sizes.}
\label{fig:perf_emr_l2}
\end{figure}
\begin{figure}
\centering
\includegraphics[width=\columnwidth]{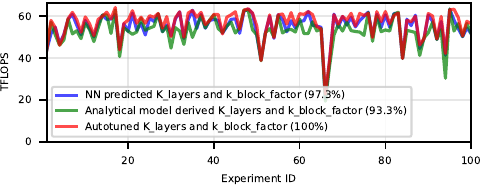}
\caption{Performance for 100 random GEMMs on EMR.}
\label{fig:gemm_models}
\end{figure}
Figure~\ref{fig:perf_emr} compares all four implementations on EMR over the 125 shapes, together with the roofline of Section~\ref{subsec:sfc_perf_model}. SFC-CA GEMM reaches a WHM of 49.05 TFLOP/s, versus 33.86 for oneDNN, 44.05 for CARMA and 27.04 for OpenBLAS, i.e.\ WHM speedups of 1.45$\times$, 1.11$\times$ and 1.81$\times$, with largest per-shape speedups of 2.64$\times$, 2.30$\times$ and 5.81$\times$, respectively. SFC-CA GEMM typically tracks the roofline within 5--25\%. It exceeds the other CA implementation (CARMA) on 120/125 shapes, and in the five remaining cases CARMA is at most 7.1\% faster. OpenBLAS is typically the slowest option.

To shed light on the source of the improvements, Figure~\ref{fig:perf_emr_l2} shows the total L2 misses of SFC-CA GEMM across all cores for two GEMMs. For $4096\times 1024\times 4096$ (Left), SFC-CA 2D (1 $C$ copy) incurs 1.34$\times$ fewer L2 misses than oneDNN. Two copies (2.5D) reduce the GEMM-phase misses (blue bars) by a further 1.18$\times$ at the cost of additional misses in the $C$ reduction (orange), and four copies (3D) by another 1.2$\times$ with a $\sim2\times$ larger reduction. In this case, the 3D variant is the fastest, with 1.61$\times$ fewer L2 misses in total and 1.52$\times$ higher performance than oneDNN (40.8 vs 26.9 TFLOP/s). For $4096\times 8192\times 4096$ (Right) the best SFC-CA GEMM uses 2 $C$ copies, reduces L2 misses by 2.9$\times$ and is 1.44$\times$ faster than oneDNN (51 vs 35.4 TFLOP/s). Thus the replication factor must balance the reduced data movement in the GEMM phase against the cost of aggregating $C$ copies, which grows with their number. We note here that not all L2 misses affect performance equally: misses on $A$/$B$ within BRGEMMs (blue bars) stall the pipeline and prevent out-of-order execution of the FMA/AMX instructions, whereas misses in streaming kernels such as the final $C$ reduction have less impact on latency.
\begin{figure*}[htbp!]
\centering
\makebox[\textwidth][c]{\includegraphics[width=1.027\textwidth,trim=0 34bp 0 9bp,clip]{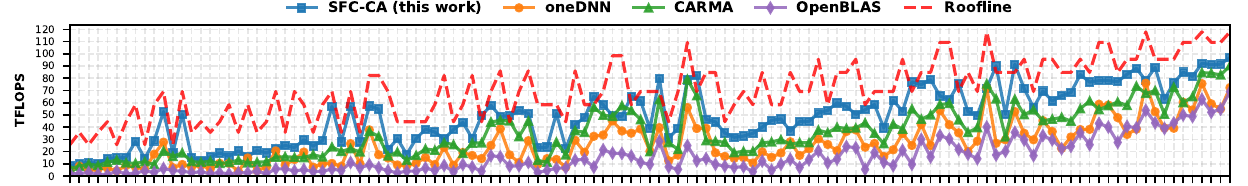}}\par\nointerlineskip
\includegraphics[width=\textwidth,trim=0 0 0 69bp,clip]{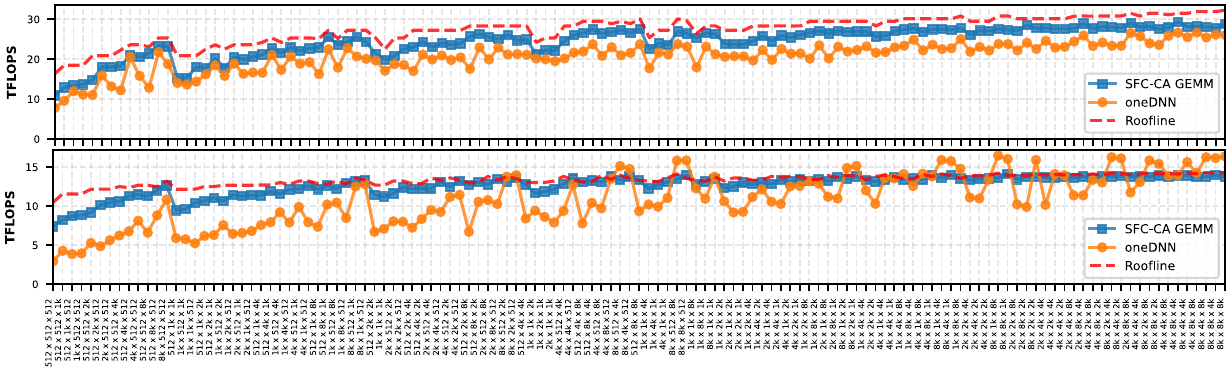}
\caption{BF16 GEMM over 125 $M\times N\times K$ shapes per platform, with a shared configuration order and x-axis labels shown only at the bottom. \textbf{Top:} GNR, comparing SFC-CA GEMM, oneDNN, CARMA, OpenBLAS and the roofline. \textbf{Middle:} ZEN5 and \textbf{Bottom:} GVT4, comparing SFC-CA GEMM, oneDNN (ACL through oneDNN on GVT4) and the roofline.}
\label{fig:perf_combined}
\end{figure*}
Figure~\ref{fig:gemm_models} shows the performance of 100 random $M\times N\times K$ GEMMs on EMR with autotuned knobs (red), knobs predicted by the Nearest-Neighbor model (blue) and by the analytical model (green) of Section~\ref{subsec:tuning}. Both models predict ($K\_layers, k\_block\_factor$) configurations very close to the optimal ones (within 3\% and 7\% respectively on average), so even without per-GEMM a priori tuning we can extract efficient ($K\_layers, k\_block\_factor$) knobs.

\subsection{Experimental results on GNR, ZEN5 and GVT4}
\label{subsec:rest_results}
Figure~\ref{fig:perf_combined}-Top shows the full GNR sweep. WHM throughput is 73.89 TFLOP/s for SFC-CA GEMM, 40.97 for oneDNN, 58.07 for CARMA and 26.22 for OpenBLAS. SFC-CA GEMM therefore improves WHM by 1.80$\times$, 1.27$\times$ and 2.82$\times$ over the three baselines, with largest per-shape gains of 5.5$\times$, 3.05$\times$ and 13.07$\times$, respectively. SFC-CA GEMM remains within 5--25\% of the roofline for the higher-operational-intensity shapes. Figure~\ref{fig:motivation} isolates the 29 highest operational intensity GNR shapes. On this subset, WHM throughput is 83.63 TFLOP/s for SFC-CA GEMM, 51.01 for oneDNN, 69.19 for CARMA and 42.82 for OpenBLAS. SFC-CA's WHM gains are 1.64$\times$ over oneDNN, 1.21$\times$ over CARMA and 1.95$\times$ over OpenBLAS.

Unlike EMR and GNR, neither ZEN5 nor GVT4 offers AMX acceleration. Their machine balance (FLOP/byte ratio) is therefore lower, GEMM performance is less sensitive to optimal data movement, and consequently the benefits of SFC-CA GEMM over the vendor libraries are expected to be smaller than on the two AMX platforms. In general, we expect that the more sensitive GEMM performance is to data movement (e.g.\ on platforms with a high compute peak via AMX), the larger the benefits of SFC-CA GEMM. CARMA is not included on ZEN5 and GVT4 because their core count (96) is not a power of two, which is a practical restriction of the CARMA algorithm.

On ZEN5 (Figure~\ref{fig:perf_combined}-Middle), SFC-CA GEMM outperforms oneDNN by up to 1.7$\times$, with a WHM of 27.7 versus 24 TFLOP/s (1.15$\times$), and typically stays within 5--10\% of the roofline. Last, Figure~\ref{fig:perf_combined}-Bottom shows the BF16 GEMM performance of SFC-CA GEMM and the Arm-optimized ACL library (via the oneDNN frontend) on GVT4. SFC-CA GEMM outperforms ACL by up to 2.5$\times$ per shape, although both achieve a WHM of 13.7 TFLOP/s, and it is within 5\% of the roofline. Considering only the first 64 shapes, which exhibit lower operational intensity, SFC-CA GEMM achieves a WHM of 13 TFLOP/s versus 10.6 TFLOP/s for ACL (1.23$\times$ speedup). Our roofline is derived by benchmarking the LIBXSMM BRGEMM to extract the $\gamma$ model parameter (see Section~\ref{subsec:sfc_perf_model}). However, in some cases the ACL GEMM microkernel outperforms the LIBXSMM one, so the ACL performance slightly exceeds the roofline. We hypothesize that ACL stores the $B$ matrix in a layout favoring the BFMMLA instruction (i.e.\ transposed $B$ matrix) and therefore achieves higher throughput, whereas LIBXSMM transposes $B$ on the fly, incurring some performance penalty in the BRGEMM.

\subsection{Case Study: Prefill Stage of LLM inference}
\label{subsec:llm_prefill}
\begin{figure*}
\centering
\includegraphics[width=0.9\textwidth]{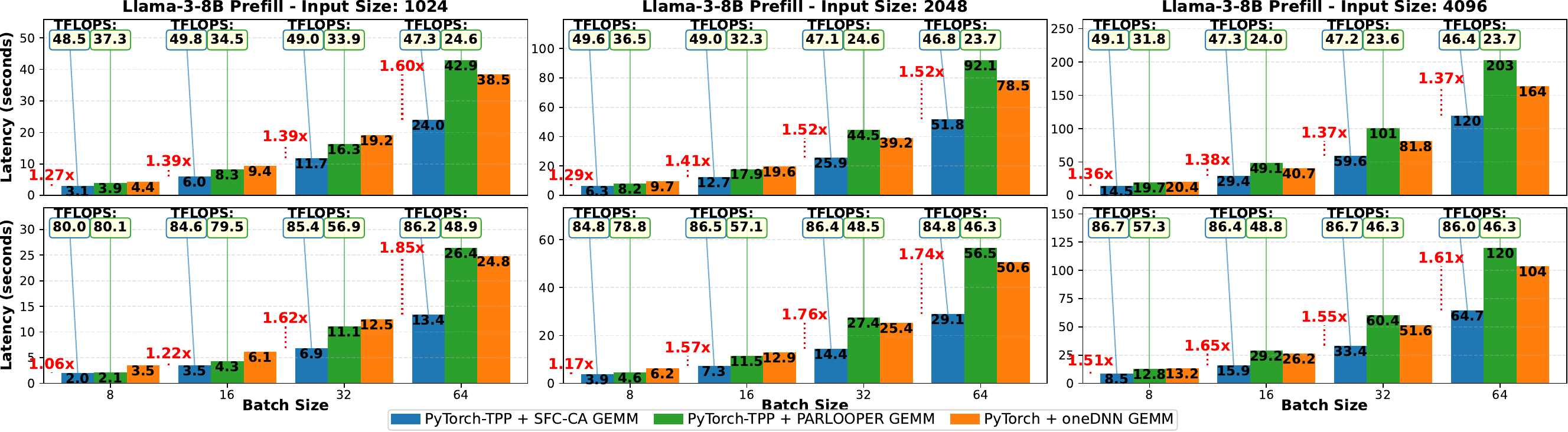}
\caption{LLM inference prefill of BF16 Llama-3 8B for various input lengths and batch sizes: i) PyTorch-TPP+SFC-CA GEMM (blue), ii) PyTorch-TPP+PARLOOPER GEMMs (green), iii) stock PyTorch+oneDNN (orange). \textbf{Top:} EMR, \textbf{Bottom:} GNR.}
\label{fig:prefill_combined}
\end{figure*}
We integrated SFC-CA GEMM into the SOTA PyTorch-TPP LLM inference framework for CPUs~\cite{georganas2021tensor,georganas2024harnessing}, replacing its default PARLOOPER-based GEMMs~\cite{georganas2024harnessing}, and evaluated the GEMM-heavy prefill of BF16 Llama-3 8B~\cite{llama8B} with input lengths 1024, 2048, 4096 and batch sizes 8, 16, 32, 64. Figure~\ref{fig:prefill_combined}-Top shows single-socket prefill latency on EMR (lower is better) for PyTorch-TPP+SFC-CA GEMM (blue), PyTorch-TPP+PARLOOPER GEMMs (green) and stock PyTorch+oneDNN (orange). SFC-CA GEMM is up to $\sim$1.6$\times$ faster than the second-best option (red labels). Within prefill (framed boxes), SFC-CA GEMM achieves 46-50 TFLOP/s versus 23-37 TFLOP/s for the PARLOOPER GEMMs, which would need further tuning. SFC-CA GEMM instead has just 2 knobs, here set by the analytical model of Section~\ref{subsec:tuning}, and delivers robust, close-to-roofline performance.

Each GNR socket consists of 3 compute dies, exposed by default as separate NUMA domains (SNC3). PyTorch-TPP therefore launches one process per domain on ``independent" sub-problems and aggregates the partial results with optimized shared-memory all-reduces, which would otherwise be unnecessary. SFC-CA GEMM instead becomes NUMA-aware \emph{implicitly}: with $K\_layers$ proportional to the number of SNC domains (e.g.\ $K\_layers=3$) it runs one ``independent" GEMM per SNC domain, followed by a final reduction. Figure~\ref{fig:prefill_combined}-Bottom shows the best variant of each setup on GNR: 1 process per socket with SFC-CA GEMM, 3 processes (1 per SNC) with PARLOOPER GEMMs, and 1 process with oneDNN. SFC-CA GEMM achieves 80-87 TFLOP/s and speedups up to $\sim$1.85$\times$ over the second-best setup. The PARLOOPER GEMMs degrade as batch size and input length grow, reaching only 46-57 TFLOP/s for (batch-size $\times$ input-size) $\geq$ 32k, which implies that tuning is required, and their all-reduces add 25--30\% to the pure GEMM time.

\subsection{Case Study: Distributed-Memory GEMM with COSMA}
\begin{figure}[t]
\centering
\includegraphics[width=0.85\columnwidth]{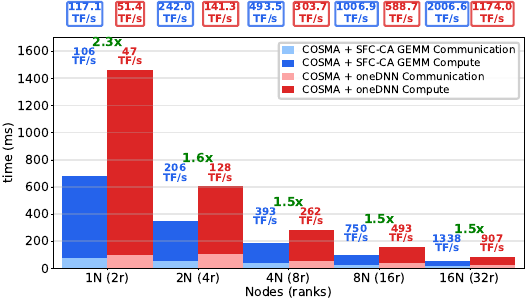}
\caption{Strong scaling of BF16 32k$\times$32k$\times$32k GEMM in COSMA with SFC-CA GEMM (blue) and oneDNN (red), on 1--16 EMR nodes (2 MPI ranks per node). Dark segments show compute, and light segments show communication plus other overhead. Boxed labels give aggregate compute-only TFLOP/s, unboxed labels effective TFLOP/s, and green labels wall-time speedups.}
\label{fig:cosma}
\end{figure}
We integrated SFC-CA GEMM as an alternative compute backend in the SOTA distributed-memory GEMM framework COSMA~\cite{kwasniewski2019red}, extended COSMA to support BF16, and strong-scaled a $32768\times32768\times32768$ GEMM on a 16-node EMR cluster (two sockets per node, one MPI rank per socket) with dual-rail ConnectX-8 NDR400 links (400\,Gb/s per rail, $\sim$94--95\,GB/s measured aggregate unidirectional node-to-node bandwidth). For each backend and node count we report the best wall-time configuration over multiple COSMA partitioning strategies.

In Figure~\ref{fig:cosma}, the end-to-end speedup over oneDNN is 2.3$\times$ on one node and 1.5--1.6$\times$ on 2 to 16 nodes. At 16 nodes, SFC-CA GEMM takes 52.57\,ms versus 77.61\,ms for oneDNN (1.34 versus 0.91 effective PFLOP/s). From 1 to 16 nodes its wall time drops from 663.56\,ms to 52.57\,ms, a 12.62$\times$ speedup and 78.9\% strong-scaling efficiency. In terms of \emph{aggregate compute-only throughput} (boxed labels on top of Figure), SFC-CA GEMM sustains $\sim$59--63 TFLOP/s per rank and reaches 2.01 PFLOP/s at 32 ranks versus 1.17 PFLOP/s for oneDNN (thus being 1.71$\times$ faster in compute). Communication (all-gather) and other overhead (light bar segments) reduce this 1.71$\times$ compute advantage to 1.5$\times$ end-to-end speedup at 16 nodes.

\section{Related Work}
\label{sec:related}
CA GEMM algorithms have been explored mostly for distributed memory systems, yielding substantial speedups~\cite{mccoll1999memory,solomonik2011communication,georganas2012communication,demmel2013communication,kwasniewski2019red,ballard2012communication,lipshitz2012communication}, while recent work~\cite{wang2025kami} introduced the CA 3D GEMM algorithm for a single GPU. CARMA's binary formulation~\cite{demmel2013communication} requires a power-of-two number of processing elements~\cite{kwasniewski2019red}, whereas the SFC-CA GEMM implementation supports arbitrary core counts.

Prior work~\cite{heinecke2008parallel,reissman2014study,chatterjee1999recursive} has also explored SFC to accelerate GEMM. However, that line of work uses SFC to derive locality-friendly tensor layouts, which fail to generalize, complicate the indexing in the GEMM microkernels and offer limited (or no) performance upside over vendor libraries. To the best of our knowledge, SFC-CA GEMM is the first GEMM algorithm that integrates the locality of SFC with communication-avoiding techniques to provably minimize data movement on the critical path and maximize throughput, portably and without specialized tensor layouts. The recursive nature of SFC yields sequential traversal schedules analogous to cache-oblivious GEMM approaches~\cite{frigo1999cache,yotov2007experimental}.

The de facto standard for SOTA GEMM performance is vendor-optimized libraries such as oneDNN~\cite{onednn} and ACL~\cite{acl}, or community-driven projects such as OpenBLAS~\cite{OpenBLAS} and BLIS~\cite{xu2023towards}. However, as we illustrated, the SOTA libraries show performance ``glass jaws", a direct artifact of the intricacies of modern platforms with matrix multiplication acceleration engines. These matrix multiplication accelerators make data movement essential to maximize performance since they skew the FLOP/byte balance of modern CPU platforms. Our work can be seamlessly integrated in GEMM libraries, alleviating such performance ``glass jaws".

Tensor compilers with auto-tuning support (e.g.~\cite{zheng2020ansor,chen2018tvm,vasilache2018tensor,kjolstad2017taco,li2024onednn}) offer an alternative to GEMM libraries. Such frameworks treat tensors as first-class objects, and provide optimizations targeting GEMMs (e.g.\ polyhedral optimizations). However, compilers struggle to optimize GEMM loop nests for increasingly complex architectures~\cite{georganas2024harnessing}. Our work is complementary, tackling the GEMM loops with SFC and CA techniques in a shape/platform-oblivious way. Due to the simplicity of SFC-CA GEMM and its platform-agnostic nature, it may serve as a robust tensor contraction tool within tensor compilers (e.g.\ recent work supporting CPU matrix multiplication ISAs~\cite{golin2024towards,thangamani2025library}), obviating the need for complicated cache-locality analysis and parallelization optimizations.

\section{Conclusions}
\label{sec:conclusions}
In this work we presented SFC-CA GEMM, a space-filling-curve-based GEMM that seamlessly implements a class of provably communication-optimal matrix multiplication algorithms. SFC-CA GEMM is compact ($\sim$30 LOC), yet it achieves SOTA results on multiple CPU platforms, outperforming vendor libraries by up to $1.8\times$ (WHM) for a range of GEMM shapes. We provide asymptotically optimal communication bounds for square and rectangular GEMM, and achieve up to 1.27$\times$ WHM speedup over the previous SOTA CA algorithm CARMA. As a compute backend, SFC-CA GEMM accelerates LLM inference prefill up to 1.85$\times$ and SOTA distributed-memory matrix multiplication up to 2.3$\times$. Although we target multi-core CPUs, the same SFC-based partitioning and tensor replication can be applied to GPU GEMMs to reduce data movement across the memory hierarchy.

\bibliographystyle{unsrt}
\clearpage
\bibliography{references}
 \scriptsize
 \noindent
 \newline Optimization Notice: Software and workloads used in
  performance tests may have been optimized for performance only on
  Intel microprocessors.  Performance tests, such as SYSmark and
  MobileMark, are measured using specific computer systems,
  components, software, operations and functions.  Any change to any
  of those factors may cause the results to vary.  You should
  consult other information and performance tests to assist you in
  fully evaluating your contemplated purchases, including the
  performance of that product when combined with other products.
  For more information go to http://www.intel.com/performance.

  \noindent Intel, Xeon, and Intel Xeon Phi are trademarks of Intel Corporation in the U.S. and/or other countries.
  
  \normalsize
\end{document}